\documentclass[aps,prl,reprint,showpacs,superscriptaddress,longbibliography]{revtex4-2}
\usepackage{mathtools,amssymb,graphicx,units}

\usepackage[plainpages=false,pdfpagelabels,colorlinks=true,linkcolor=red,urlcolor=blue,citecolor=blue,pdftitle={Title},pdfauthor={},pdfdisplaydoctitle=true,pdfduplex=DuplexFlipLongEdge]{hyperref}
\usepackage{verbatim}
\usepackage[english]{babel}
\usepackage{color}
\usepackage{ulem}

\begin{document}
\title{Many-body Localization in Clean Chains with Long-Range Interactions}

\author{Chen Cheng}
\email{chengchen@lzu.edu.cn}
\affiliation{Key Laboratory of Quantum Theory and Applications of MoE $\&$ Lanzhou Center for Theoretical Physics $\&$ Key Laboratory of Theoretical Physics of Gansu Province, Lanzhou University, Lanzhou, Gansu 730000, China}

\begin{abstract}
The strong long-range interaction leads to localization in the closed quantum system without disorders. Employing the exact diagonalization method, the author numerically investigates thermalization and many-body localization in translational invariant quantum chains with finite Coulomb interactions. In the computational basis, excluding all trivial degeneracies, the interaction-induced localization is well demonstrated in aspects of level statistics, eigenstate expectation values, and the Anderson localization on graphs constructed of the many-body basis. The nature of localization for generic eigenstates is attributed to the quasi-disorder from the power-law interactions. However, due to the real-space symmetries, the long-time dynamics is dominated by the degenerate eigenstates and eventually reach homogeneity in real space. On the other hand, the entanglement entropy exhibits the size-dependence beyond the area law for the same reason, even deep in the localized state, indicating an incomplete localization in real space.
\end{abstract}
\maketitle

\paragraph{Introduction.---}

Dynamics of the isolated quantum many-body systems have been extensively studied in recent years, and a new paradigm focus on the out-of-equilibrium quantum phase transition has been established. In the interacting and nonintegrable setting, the own unitary dynamics of a generic quantum system drives itself to thermalized states, and this ergodic phase is characterized by the eigenstate thermalization hypothesis (ETH)~\cite{Deutsch1991,Srednicki1994,Rigol2008,Alessio2016}. In contrast, the system affected by sufficient disorders features many-body localization (MBL) and fails to thermalize. Its long-time evolution partially preserves the encoded information of the initial state~\cite{Nandkishore2015,MBL_RMP2019}. Disorder-induced thermal-MBL transition can be characterized by essential differences in aspects such as level statistics~\cite{Oganesyan2007,Pal2010,Atas2013}, entanglement entropy of eigenstates and its real-time spread behaviours~\cite{Znidaric2008,Bardarson2012,Serbyn2013}, and have attracted many experimental probes in various platforms~\cite{Schreiber2015,Choi2016,Smith2016,Luschen2017,Roushan2017,Xu2018,Wei2018,Rispoli2019,Guo2020}.  

Although MBL is mainly studied in the disordered systems as it roots on the non-ergodicity of the well-known disorder-induced Anderson localization for noniteracting particles~\cite{Anderson1958,EverRMP2008}, disorder-free MBL and other ETH breakdown scenarios in the absence of disorder have attracted increasing interests~\cite{Smith2017,Schiulaz2015,Schulz2019,Doggen2021,Guo2021,Doggen2021,Li2020,Barbiero2015,Li2021,Korbmacher2023,Khemani2020,Herviou2021,Frey2022,Chakraborty2022,Moudgalyabook2019,MoudgalyaPRX2022,Sala2020,Moudgalya2022}. The related studies brings in attentions to new issues like Stark-MBL~\cite{Schulz2019,Doggen2021,Guo2021}, quantum many-body scars~\cite{Turner2018,Bluvstein2021,Moudgalya2022} and Hilbert space fragmentation~\cite{Khemani2020,Sala2020,Herviou2021,Mukherjee2021,Frey2022,Moudgalya2022}, and involves systems with conserved dipolar momentum~\cite{Sala2020}, pairing hoppings~\cite{Herviou2021,MoudgalyaPRX2022} and strong constraints~\cite{Nandkishore2019,Tomasi2019}, to name a few. Specifically, the one-dimensional system with short-range interactions can be projected to a dynamically constrained model in the strongly interacting limit, where its Hilbert space fragments into an exponential number of disconnected sectors~\cite{Tomasi2019,Rakovszky2020}. The dynamics in different sectors can be different, and some sectors feature complete localization with any finite disorders~\cite{Tomasi2019}.
 
In parallel, the effect of long-range couplings on the context of thermalization and localization attracts interest in many aspects~\cite{Singh2017,Burin2015,Tikhonov2018,Maksymov2020,Bhakuni2020,Deng2018,Deng2019,Wei2019,Nosov2019,Nosov2019_prb99224208,Modak2020,Modak2020prr,Lerose2020,Deng2020,Kuwahara2021,Vu2022,Marie2022,Lukin2022,Yousefjani2023,Huang2023}. While both long-range interaction and hopping are generally believed to allow far-distance quantum correlations and build up faster thermalization~\cite{Burin2015,Tikhonov2018,Maksymov2020,Bhakuni2020}, it is reported that long-range couplings can lead to anomalous algebraic localization~\cite{Deng2018,Deng2019} and stabilize Stark-MBL~\cite{Lukin2022}. Moreover, the logarithmic growth of the entanglement entropy over time, known as a unique characteristic of the many-body localization, is also found in the noninteracting system with long-range hoppings~\cite{Singh2017}. In recent studies, the long-range interactions, with sufficient but finite strength, are found to result in Hilbert space fragmentation and localization in the absence of disorders~\cite{Barbiero2015,Khemani2020,Li2021,Korbmacher2023}. However, the real-time evolution in this disorder-free MBL phase features abnormal two-stage dynamics: it has MBL-like behavior in a considerable time range but eventually thermalizes in the long-time limit. The latter is attributed to higher-order virtual excursions between different fragmented Hilbert sections~\cite{Li2021}. 

This coexistence of localization and thermalization in the clean long-range interacted system is beyond the picture of disorder-induced MBL. In this work, the author revisits the disorder-free localization induced by long-range interactions and aims to understand better this incomplete MBL phenomenon. The interests of long-range interacted systems also rely on the widely existing power-law interactions between particles in nature, and the experimental developments simulating long-range interactions, in the platforms of ultra-cold polar molecules~\cite{Gorshkov2011,Gorshkov2011a}, trapped ions~\cite{Richerme2014,Jurcevic2014} and Rydberg gases~\cite{Bernien2017}.

\paragraph{Model and method.---}

The present work focuses on a one-dimensional bosonic Hubbard Hamiltonian with power-law interactions 
\begin{align}
    \hat{\cal H} = -t\sum_{\langle i j\rangle}\left( \hat{b}_i^\dagger \hat{b}_{j}  + H.c.\right) + V\sum_{i<j}R_{i,j}^{-\beta}\hat{n}_i \hat{n}_j,
    \label{eq:ham}
\end{align}
where $\hat b^\dagger_i$ ($\hat b_i$) is the hard-core bosonic creation (annihilation) operator at site $i$ and $\hat n_i=\hat b^\dagger_i\hat b_i$ is the density operator; $R_{i,j}$ denotes the distance between sites $i$ and $j$ in the system with size $L$. The strong long-range interaction leads to a Mott insulating groundstate of Eq.~\eqref{eq:ham}, and supports interesting groundstate physics when both charge and spin degrees of freedom are considered~\cite{Cheng2015} or in higher dimensions~\cite{Yang2022,Gong2023}. Regarding the high energy states and dynamics, it is recently reported that sufficient dipole interactions result in Hilbert space fragmentation and MBL~\cite{Li2021}.

The author considers the Coulomb interaction ($\beta=1$), which is longer ranged than the dipole form and eliminate multi-band structure at relatively large $V$ in the density of states (DOS) to have better level statistics. The Hamiltonian~\eqref{eq:ham} conserves the total particle number $N= \langle \hat{N}\rangle = \sum_i \langle {\hat n}_i\rangle$, and the present work is restricted to the largest sector with half-filling $N=L/2$. The periodic boundary conditions are adopted to minimize the boundary effect. To rule out all trivial degeneracies and reduce the computational cost of the full exact diagonalization, the Hamiltonian matrix is constructed in basis $|s\rangle$ preserving the following symmetries as~\cite{Sandvik2010},
\begin{align}
    \hat{N}\hat{T}\hat{P}\hat{Z}|s\rangle = N e^{\i k} pz|s\rangle.
    \label{eq:basis}
\end{align}  
Here $\hat T$ is the momentum operator and $k=2m\pi/L$ ($m=-L/2,\cdots,0,\cdots,L/2$) is the momentum; $\hat P$ ($\hat Z$) represents the reflection (particle-hole inversion) operator and $p=\pm 1$ ($z=\pm 1$) is the corresponding parity. The Hilbert space dimension ${\cal N}$ is of the order of $\propto 2^L /(L \times L \times2 \times2)$, which is about $10^5$ for the largest system size $L=26$ investigated. For all numerical results, $t=1$ is set as the energy scale.

\paragraph{Level statistics and eigenstate expectations.---}

\begin{figure}[!b]
    \includegraphics[width=1\columnwidth]{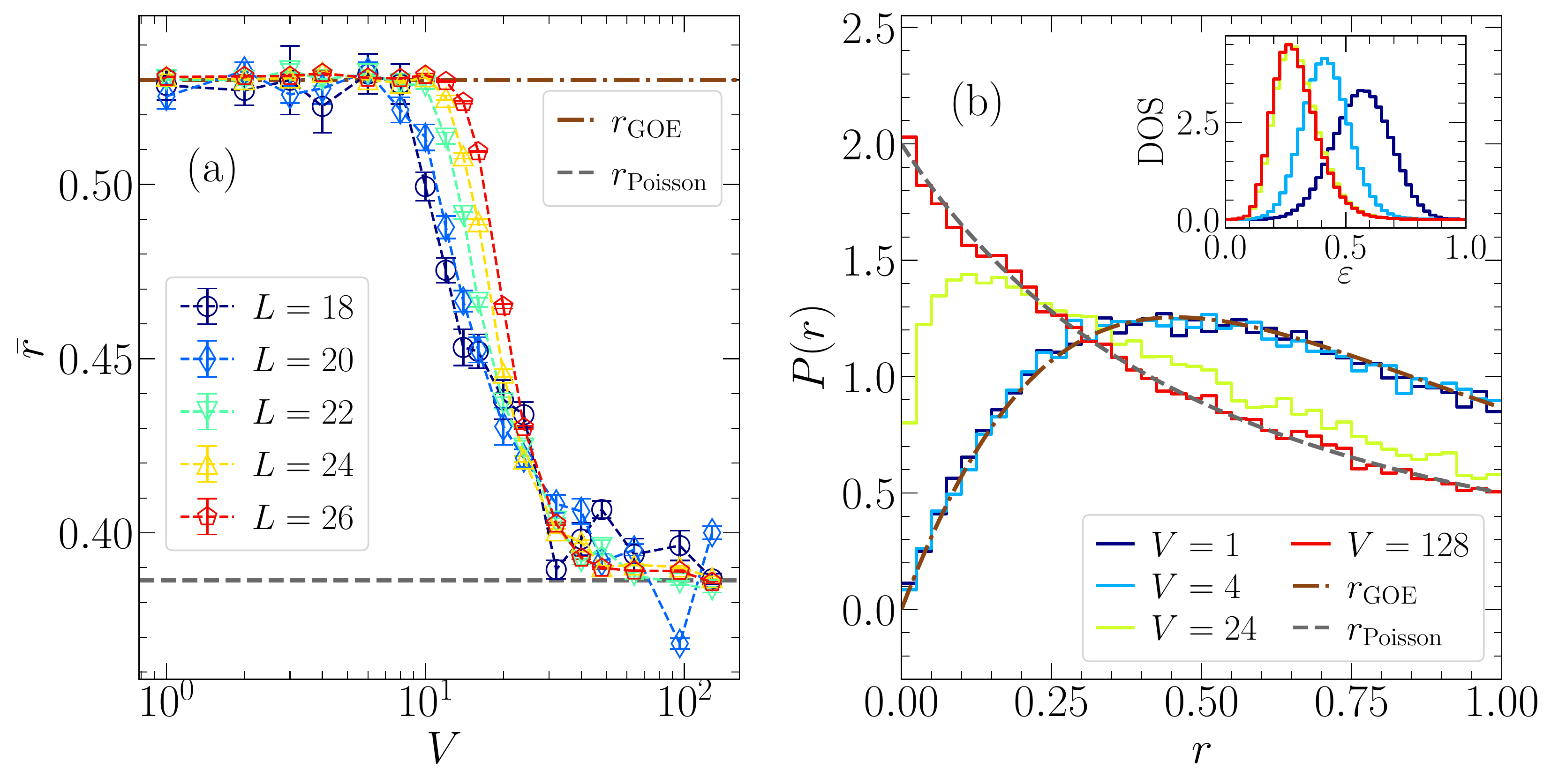}
    \caption{(a) The mean ratio of adjacent gaps $\bar{r}$ versus interaction strength $V$ for different system size $L$. Here the mean value and error are from the average over eight sectors with conserved momentum ($k=0,\pi/2$) and parities ($p\pm1$, $z \pm 1$); within each sector, we consider energy levels in the middle half of the spectrum. (b) The distribution $P(r)$ for $L=26$ and different $V$ in the sector $\{k=0,p=z=1\}$. The Inset displays the density of state in the same sector as a function of energy density $\varepsilon = (E-E_{\min})/(E_{\max} - E_{\min})$.}
    \label{fig:roag_dos}
\end{figure}

Thermalization and localization are generally related to the ergodic and nonergodic regimes, which can be characterized by the spectrum structure and specifically quantified using the ratio of adjacent gaps~\cite{Oganesyan2007,Pal2010,Atas2013}, $r_{\alpha}\equiv \min(\delta_{\alpha+1},\delta_\alpha)/\max(\delta_{\alpha+1},\delta_\alpha)$, with $\delta_\alpha=E_\alpha-E_{\alpha-1}$ is the gap between adjacent energy levels and $E_\alpha$ is the $\alpha$th eigenenergy. As displayed in Fig.~\ref{fig:roag_dos}(a), the average $\bar r$ agrees with the value predicted by the Gaussian orthogonal ensemble (GOE) at small interactions, indicating the ergodic phase in this region. At large $V$s, $\bar r$ approaches the value that obeys the Poisson distribution and shows localization characteristics. The results from different system sizes suggest an ergodic-nonergodic transition as the jump of $\bar r$ becomes sharper as $L$ increases. 

The ergodic and non-ergodic phase is further confirmed by the explicit distribution of $r$ for several typical interactions at the largest $L=26$, as shown in Fig.~\ref{fig:roag_dos}(b). The DOS in the inset of (b) significantly differs at different $V$s; therefore, the author takes the average over the middle half of the spectrum to capture the physics at the infinite temperature instead of focusing on the fixed energy density. The continuous band manifests no obvious spectrum fragmentation at a large interaction $V=128$, which differs from the short-range and dipole interactions cases. Note that in the limit $V\rightarrow \infty$, it is obvious that the eigenenergies approach the spectrum of the basis~\eqref{eq:basis} with banded DOS and massive degeneracies, which is not the interest of the present work.

\begin{figure}[!t]
    \includegraphics[width=1\columnwidth]{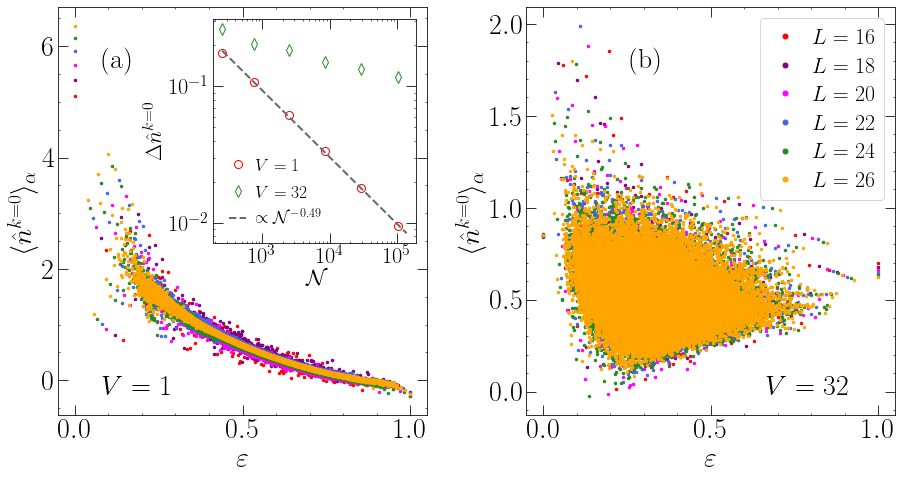}
    \caption{Eigenstate expectation values $\langle \hat{n}^{k=0} \rangle_\alpha$ for (a) $V=1$ and (b) $V=32$ and different $L$. The Inset of the panel (a) shows the fluctuations of consecutive eigenstate expectation values $\Delta n^{k=0}$ as a function of the Hilbert space dimension $\cal N$, where the dashed line displays the fitting to ${\cal N}^\gamma$ with $\gamma=-0.49$. Here the results correspond to the sector $\{k=0,p=z=1\}$.}
    \label{fig:eev}
\end{figure}

In the thermalized phase, ETH predicts that the eigenstate expectation value of a generic few-body operator is a smooth function of energy~\cite{Deutsch1991,Srednicki1994,Rigol2008,Alessio2016}. On the contrary, in the MBL phase, the physical quantity is discontinuous even for eigenstates with very close energies. We verify the ergodicity of the system along this line by examining the zero momentum occupancy $\hat{n}^{k=0} = \frac{1}{L}\sum_{i,j}\hat{b}_i^\dagger \hat{b}_j$ as well as the corresponding eigenstate-to-eigenstate fluctuation $\Delta \hat{n}^{k=0} \equiv |\langle \hat{n}^{k=0} \rangle_\alpha - \langle \hat{n}^{k=0}\rangle_{\alpha+1}| $, where $\langle \rangle_\alpha$ denotes the expectation value of eigenstate $|\alpha\rangle$. As displayed in Fig.~\ref{fig:eev}(a), for the small $V=1$, the possessed region of $\langle \hat{n}^{k=0} \rangle$ becomes narrower as the system size increases, which implies a smooth function of energy in the thermodynamic limit. In contrast, the same quantity in the MBL phase ($V=32$) in panel (b) shows little system size dependency. In the inset of Fig.~\ref{fig:eev} (a), there is a clear power-law behavior of the $\Delta \hat{n}^{k=0}$ as a function of Hilbert space dimension $\cal N$ 
in the thermal phase, where the fitting exponent $-0.49$ is in consistence with the ${\cal N}^{-1/2}$ behavior predicted by ETH analysis~\cite{Beugeling2014}. 

\paragraph{Anderson transition in random graphs.---}

Anderson localization of single-particle orbitals is known to display multifractality at criticality~\cite{EverRMP2008}. Recent studies have observed a more generic Hilbert space multifractality in interacting systems and employ it to describe thermal-MBL transitions~\cite{Backer2019,Mace2019,Pietracaprina2021,Wang2021,Sutradhar2022}. Rewriting the Hamiltonian~\eqref{eq:ham} in the computational basis $|s\rangle$ [see Eq.~(\ref{eq:basis})] as 
\begin{align}
    {\cal H} = \sum_s \mu_s |s\rangle\langle s| + \sum_{s \neq s^\prime} t_{s s^\prime}|s \rangle\langle s^\prime|,
    \label{eq:ham_al}
\end{align}
the Thermal-MBL transition degenerates to a generalized Anderson localization problem, and the many-body basis $|s\rangle$ can be considered as complex Anderson orbitals, where the summation of Coulomb bonds provides $\propto {\cal N}$ different values for diagonal elements and $\mu_s$ acts like the quasi-disorder. For arbitrary many-body wavefunction $|\Psi\rangle = \sum_s \psi_s|s\rangle$, the interests related to multifractality lie in the $m$th participation entropy defined as
\begin{align}
    S_\lambda = \frac{1}{1-\lambda}\ln \left( \sum^{\cal N}_{s=1} |\psi_s|^{2\lambda} \right).
    \label{eq:sq}
\end{align}
Specifically, $S_1=-\sum_s |\psi_s|^2 \ln |\psi_s|^2$ is the Shannon entropy, and $S_2$ is the logarithm of the inverse participation ratio (IPR)~\cite{VISSCHER1972477} defined as $\mathrm{IPR}=1/\sum_s|\psi_s|^4$. For a finite system with Hilbert space dimension $\cal N$, the $\lambda$-dependent fractal dimension ${\cal D}_\lambda$ measures the ergodic fraction of the Hilbert space, with $S_\lambda({\cal N})={\cal D}({\cal N})\ln{\cal N}$. The asymptotic scaling approaching the infinite dimension follows the form ${\cal D}_\lambda({\cal N}) = {\cal D}_\lambda^\infty - f_\lambda$ with a non-universal constant $f_\lambda$~\cite{Backer2019}, therefore one can extract ${\cal D}_\lambda^\infty$ by the linear fitting $S_\lambda({\cal N}) = {\cal D}_\lambda^\infty \ln{\cal N} -f_\lambda$. 

\begin{figure}[!t]
    \includegraphics[width=1\columnwidth]{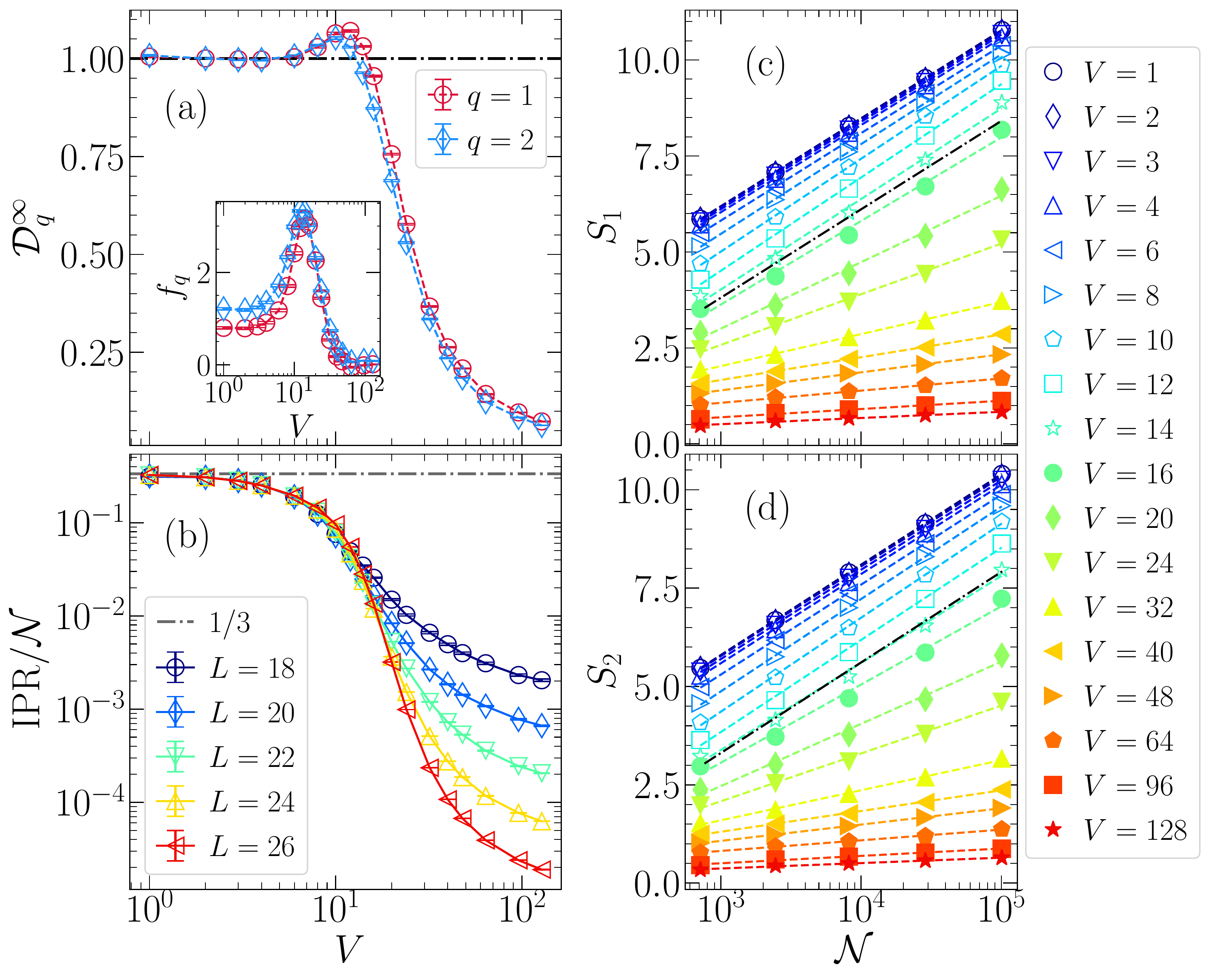}
    \caption{(a) The Multifractal dimension ${\cal D}_\lambda^\infty$ as a function of $V$, and the fitting parameter $f_\lambda$ in the inset. (b) $\mathrm{IPR}/{\cal N}$ as a function of $V$, where the horizontal line marks $1/3$ from ETH prediction. In (a) and (b), the mean value and error are from the average over eight sectors with conserved momentum $k$ ($0,\pi/2$), parities $p$ ($\pm1$) and $z$ ($\pm 1$). (c) and (d) displays the linear fitting ${\cal D}_q({\cal N}) = {\cal D}_q^\infty - f_q$ for $S_1$ and $S_2$, respectively. In (c) and (d), the data are from the sector $\{k=0,p=z=1\}$, and the black dashed line has a slope of $1$. }
    \label{fig:multifrac_dim}
\end{figure}

As displayed in Fig.~\ref{fig:multifrac_dim}(a), ${\cal D}_\lambda^\infty$ as a function of interaction well captures the dynamic transition: at small $V$s, ${\cal D}_\lambda^\infty=1$ indicates fully ergodicity in the thermal phase; at larger interactions, ${\cal D}_\lambda^\infty<1$ shows the multifractal behavior in the MBL phase. The fitting parameter $f_\lambda$ in the inset also sharply peaks at the transition point. Note that $f_\lambda$ is always positive and approaches zero deep in the MBL region, which differs from the disorder-induced thermal-MBL transition where the fitting parameter changes its sign at the critical point~\cite{Mace2019}. Fig.~\ref{fig:multifrac_dim}(c) and (d) explicitly display the fittings of $S_1$ and $S_2$, in the sector $\{k=0,p=z=1\}$ as an example. Both of them provide the critical point around $14$, depicted by the black dash-dotted line with slope 1. Interestingly, for all $V$s and $L$s visited, $S_\lambda$ versus $\cal N$ shows nice linear behaviors, which also differs from the disordered case. 

The ergodic-nonergodic transition can also be characterized by displaying $\mathrm{IPR}/{\cal N}$ for different system sizes and interactions, as shown in Fig.~\ref{fig:multifrac_dim}(b). In the thermal region, $\mathrm{IPR}/{\cal N}$ is almost independent of $L$, and the value agrees with the GOE prediction $\mathrm{IPR}={\cal N}/3$ at very small interactions. In the localized region, $\mathrm{IPR}/{\cal N}$ decays rapidly as $L$ increases. From Fig.~\ref{fig:multifrac_dim}(b-d), a generalized volume-law participate entropy for the thermal state can be observed in the Hilbert space. 

\paragraph{Real-time dynamics.---}

\begin{figure}[!t]
    \includegraphics[width=1\columnwidth]{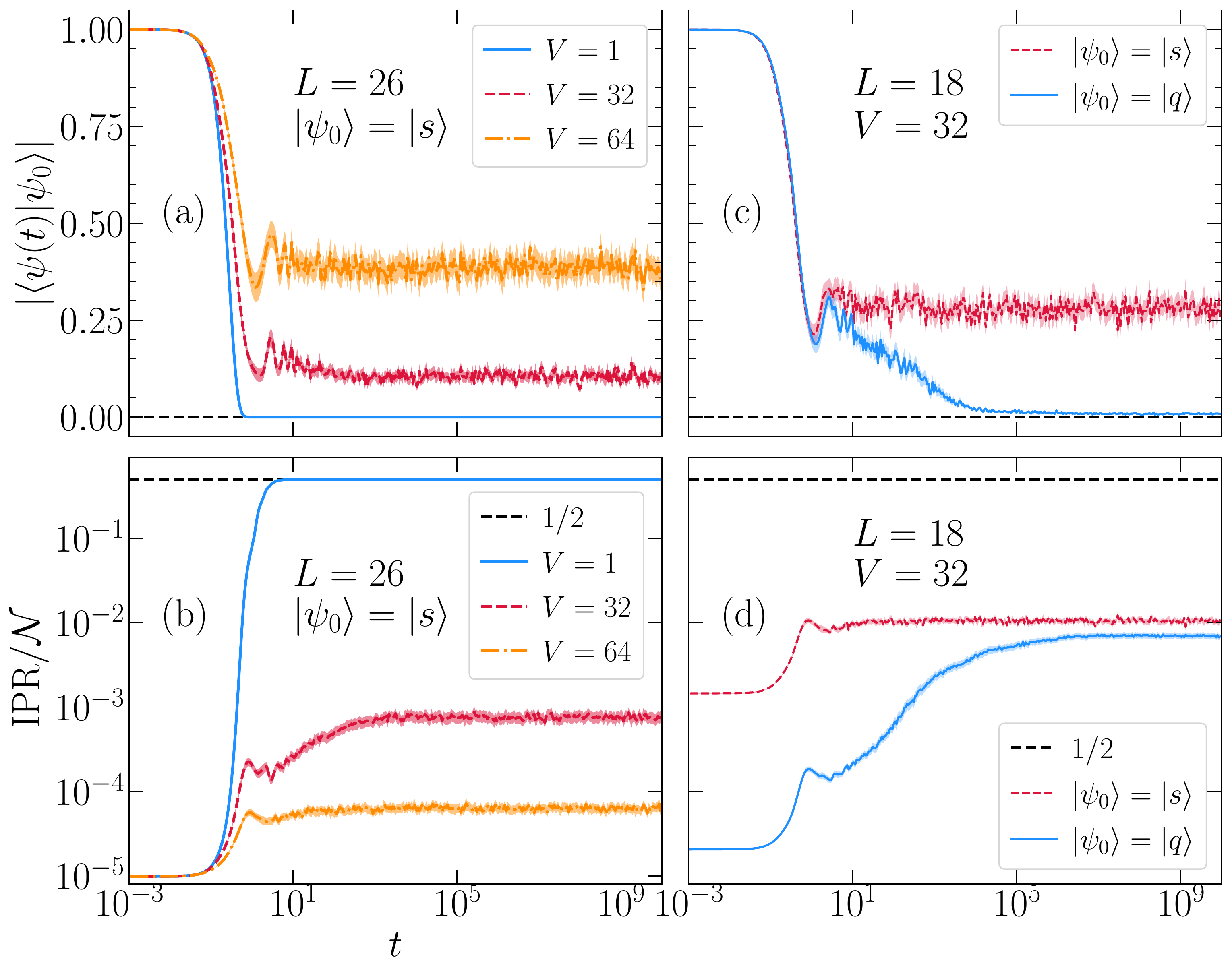}
    \caption{Real-time evolution of (a) fidelity $|\langle \psi(t)|\psi(0) \rangle|$ and (b) $\mathrm{IPR}/{\cal N}$ starting from initial state $|s\rangle$ maintaining all symmetries in the sector $\{k=0,p=z=1\}$, for $L=26$ and different $V$s. Panels (c) and (d) compare the evolution from product state $|q\rangle$ and the corresponding symmetric states $|s\rangle$. The evolution has been averaged over 64 random initial states in the middle half of the initial state spectrum.}
    \label{fig:evo}
\end{figure}

Localization in Hilbert space can be further confirmed by checking the real-time dynamics starting from a random state in the computational basis $|s\rangle$. In Fig.~\ref{fig:evo}(a), the time-evolved fidelity $|\langle \psi(t)|\psi(0) \rangle|$ decays rapidly to zero for $V=1$, since the dynamical return probability of a single site in the Hilbert space is infinitesimal in the chaotic system. In contrast, the Hilbert space can not be fully visited in the localized phase, and the fidelity keeps a finite value at large interactions ($V=32,64$). Unlike the two-step dynamics for the homogeneity parameter observed in Ref.~\cite{Li2021}, this high overlap with the initial state shows no indication for further decay, at least at the longest time (above $10^{10}$) visited in this work. The evolution of IPR features similar dynamics. As displayed in Fig.~\ref{fig:evo}(b), ${\mathrm {IPR}}/{\cal N}$ approaches the zero in a relatively short time in the thermal phase, but fluctuates around small values in the localized state. Here all the real-time evolution is averaged over 64 randomly initial states in the middle half of the spectrum to mimic the same physics with static quantities such as the level statics~\footnote{By the averaging over many random initial states, we neglect the influence of possible low-measure (or even measure-zero) special initial states, like one-mover blocks in Ref.~\cite{Tomasi2019} and the localization plateau for the special initial state in Ref.~\cite{Li2021}. The interest of this work lies in the general strong ETH break scenario, but not the Hilbert space fragmentation or special state localization such as quantum scars. }.

According to the above analysis on level statistics, eigenstate expectation values, Anderson localization in many-body Hilbert space, and real-time dynamics of fidelity and IPR, increasing Coulomb interaction seems to induce a standard thermal-MBL transition, similar to the disorder case. This conclusion, however, disagrees with the results in boson chains with dipole interactions, where the eventual thermalization is observed in the MBL phase by two-stage dynamics at long times~\cite{Li2021}. To understand this inconsistency, the author reinvestigates the localized phase in the basis of products $|q\rangle$, which conserves only the total particle $N$~\footnote{The product state and the symmetric correspondence follow the relation $|s\rangle=1/(\sqrt{2\times2\times L})\sum_r (1+p\hat{P})(1+z\hat{Z})T^r|q\rangle$. The latter is the basis set in the sectors with $k=0$/$\pi$. In this work, the author focuses on the evolution with $k=0,p=z=1$. }. As shown in Fig.~\ref{fig:evo}(c), the dynamics of different kinds of initial states in the localized phase have remarkable differences. For the evolution starting from a product state, the fidelity features a slow logarithmic dynamic at large times and eventually approaches zero, providing clues of a final thermalization. However, although $\mathrm{IPR}$ of the product state $|q\rangle$ continues to increase at a longer time range than the symmetric state $|s\rangle$, the two curves almost merge at the long-time limit and show the lack of ergodicity in the full Hilbert space. The different indications from fidelity and IPR may arise from the degeneracy of the system in the product basis, in which case the evolution fully visits the subspace spanned by the degenerated eigenstates (with a dimension proportional to $L$) but not the full Hilbert space.

\paragraph{Entanglement entropy.---}

\begin{figure}[!t]
    \includegraphics[width=1\columnwidth]{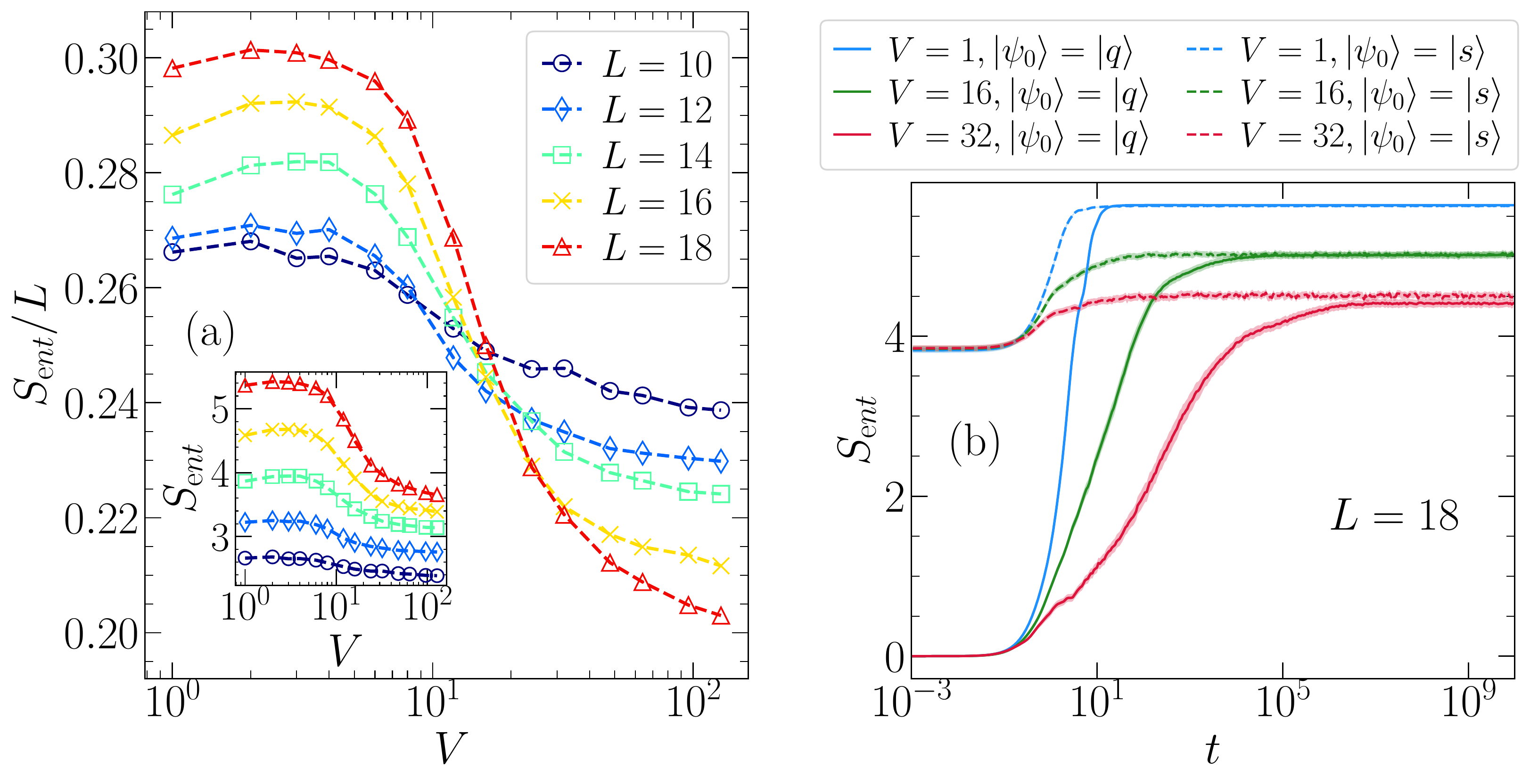}
    \caption{ (a) The entanglement entropy per site $S_\mathrm{ent}/L$ of the eigenstates versus $V$ for different system sizes; Inset displays $S_\mathrm{ent}$. (b) Real-time evolution of $S_\mathrm{ent}$ for $L=18$ and three typical $V$s, compared between the evolution from product state $|q\rangle$ and the corresponding symmetric states $|s\rangle$. 
    The choice of initial states is the same as Fig.~\ref{fig:evo}.}
    \label{fig:Sent}
\end{figure}

The eigenstate entanglement entropy, which features volume/area law in the thermal/MBL phase, is widely used to characterize the thermal-MBL transition. The half-chain entanglement entropy $S_\mathrm{ent}= \mathrm{Tr} \rho_A \ln \rho_A$ with $A$ containing the first half of sites is discussed in the following.  In the inset of Fig.~\ref{fig:Sent}(a), for the systems investigated, $S_\mathrm{ent}$ increases as $L$ increases, even in the localized state with large interactions. Furthermore, by checking the entanglement entropy per site, the information of the thermal-MBL transition can be extracted from the crossings of $S_\mathrm{ent}/L$ versus $V$ for different system sizes. Regardless that one can not rule out the finite size effect, the current data for system sizes up to $L=18$ shows an entanglement entropy behavior beyond volume law in the thermal phase, and between area and volume law at large interactions. Although it is hard to properly extrapolate $S_\mathrm{ent}$ to the thermodynamic limit with current results, one can estimate minimum size dependence of a generic translational invariant state as $S_\mathrm{ent} \propto \ln L$ in the localized phase.  

The evolution of entanglement entropy is displayed in Fig.~\ref{fig:Sent}(b). The symmetric initial state $|s\rangle$ is already highly entangled in real space, and the evolution starting from which reaches the long-time limit in a relatively short time, despite the interacting strength. For the non-entangled product states, the quench dynamics is different in different phases: $S_\mathrm{ent}$ rapidly saturates in the thermal phase ($V=1$) but has a long-time slow growth in the localized phase ($V=32$). Combining the evolutions in Fig.~\ref{fig:evo}(d) and~\ref{fig:Sent}(b), the evolution starting from a product $|q\rangle$ may have similar short-time relaxation nature as its symmetric correspondence $|s\rangle$, mostly contributed by the direct hoppings of the Hamiltonian~\eqref{eq:ham}. At long times, higher-order processes, especially for those between product states generated by symmetry operators from the same $|q\rangle$, dominate the dynamics and eventually reach ergodicity in a subset of the Hilbert space. This subspace, which is determined by the long-time evolution of the corresponding symmetric states $|s\rangle$, can be a small fraction of the full Hilbert space at large interactions. 

\paragraph{Conclusion and discussion.---}

The strong Coulomb interaction in the one-dimensional bosonic chains leads to localization and ETH breakdown without disorders. In the computational basis ruling out all trivial symmetries, interaction-induced localization transition is nicely characterized by the level statistics, eigenstate expectation values, multifractality in the Hilbert space, and the time evolution of fidelity and IPR. Here the nature of localization is attributed to the quasi-disorder provided by the Coulomb interactions, and the localization for generic eigenstates is well-defined in Hilbert space without borrowing the concept of Hilbert space fragmentation. It is worth noting that the interaction is relatively strong but still finite with a non-banded DOS, away from the strong-interacting limit. 

However, due to real space symmetries and the consequent degeneracy, the real-time dynamics of a product state in the localized phase exhibits two different time scales. After a short-time relaxation on account of the direct hopping term, the higher-order processes dominate the long-time behaviors, ultimately merging the evolution of the corresponding symmetric state. On the other hand, the entanglement entropy for the localized eigenstate with the real-space symmetries has a size dependence beyond the area law. This long-range interaction-induced localization in clean chains is beyond the picture of the $l$-bits~\cite{Imbrie2017,Rademaker2017} of the disorder-induced MBL and deserves further investigation.

\paragraph{Acknowledgements.---}
This work is supported by the National Natural Science Foundation of China (grant nos.~12174167, 11904145, 12247101), the 111 Project under Grant No.~20063, and the Fundamental Research Funds for the Central Universities.

\bibliography{mbl_lr_int}

\end{document}